\documentclass[aps, pra, superscriptaddress, onecolumn, floatfix, showpacs, 10pt,notitlepage]{revtex4-1}
\pdfoutput=1
\usepackage{hyperref}
\usepackage{graphicx}
\usepackage{epstopdf}
\usepackage{subfigure}

\usepackage{amsmath}

\newcommand{\bs}{\boldsymbol}

\begin{document}

\title{Conical intersections for light and matter waves}

\author{Daniel Leykam}
\affiliation{Division of Physics and Applied Physics, School of Physical and Mathematical Sciences, Nanyang Technological University, Singapore 637371, Singapore}

\author{Anton S. Desyatnikov}
\affiliation{Physics Department, School of Science and Technology, Nazarbayev University, 53 Kabanbay Batyr Ave., Astana, Kazakhstan}
\affiliation{Nonlinear Physics Centre, Research School of Physics and Engineering, The Australian National University, Canberra ACT 2601, Australia}

\begin{abstract}
We review the design, theory, and applications of two dimensional periodic lattices hosting conical intersections in their energy-momentum spectrum. The best known example is the Dirac cone, where propagation is governed by an effective Dirac equation, with electron spin replaced by a ``fermionic'' half-integer pseudospin. However, in many systems such as metamaterials, modal symmetries result in the formation of higher order conical intersections with integer or ``bosonic'' pseudospin. The ability to engineer lattices with these qualitatively different singular dispersion relations opens up many applications, including superior slab lasers, generation of orbital angular momentum, zero-index metamaterials, and quantum simulation of exotic phases of relativistic matter.
\end{abstract}


\maketitle

\section{Introduction}

A conical intersection occurs when two or more energy surfaces in a parameter space become degenerate at a point. Such a singular degeneracy signifies the breakdown of the simplest single mode or scalar approximation, demanding the introduction of concepts such as geometric phases and pseudospinor or vector order parameters, leading to phenomena that are not just quantitatively, but also qualitatively different from the scalar case~\cite{condensed_review}. Originally considered pathological, requiring high symmetries absent from any realistic system~\cite{jahn-teller}, we now recognise the ubiquity of conical intersections, appearing across diverse disciplines ranging from chemical reaction dynamics~\cite{chemical_review} to condensed matter systems such as graphene~\cite{graphene_review} and more recently to the closely related fields of photonics~\cite{dennis2009,lu2014} and matter waves~\cite{bec1,bec2}.

The focus of the present review is on conical intersections in two-dimensional periodic media and their application to photonic and matter waves. In this case the parameter space is the in-plane wavevector (Bloch momentum), and the energy surfaces are the Bloch bands of the lattice. Notably in this situation conical intersections lead to wave propagation that emulates aspects of \emph{relativistic} wave physics previously deemed hideously impractical to achieve, but now routinely observed in tabletop~\cite{klein_tunnelling} or even on-chip experiments~\cite{on_chip} and even finding useful applications.

The recent tremendous interest in these conical intersections derives from graphene, a two-dimensional honeycomb lattice of carbon atoms hosting the simplest type of conical intersection, ``Dirac cones,'' in its electronic band structure~\cite{graphene_review,klein_tunnelling}. Other branches of physics including photonics have drawn on this discovery to engineer so-called ``artificial graphenes'' displaying analogous properties~\cite{polini2013}. The aim of this review is to highlight how the interesting physics of conical intersections is not just limited to emulating effects observed in graphene; in fact the ability to coherently control wavepackets and potential landscapes for photonic and matter waves leads to novel effects not readily accessible or with no analogue in ``real'' graphene. We have seen a rapid growth of literature on this topic, with cross-disciplinary applications not limited to the electronic properties of graphene's Dirac cones.

We begin by briefly surveying periodic media displaying conical intersections in Sec.~\ref{sec:examples}, with focus on photonic lattices, crystals, and metamaterials. Sec.~\ref{sec:theory} outlines their theoretical description, highlighting qualitative differences between the ``fermionic'' Dirac cone and higher-order ``bosonic'' intersections associated with half-integer and integer pseudospins respectively. We explore some practical applications of photonic conical intersections in Sec.~\ref{sec:applications}, before concluding in Sec.~\ref{sec:conclusion} with a discussion of links to other fields and some interesting open problems. 

\section{Design}
\label{sec:examples}

Perhaps the first example of wave propagation governed by a conical intersection was discovered by Hamilton in 1832: conical diffraction in biaxial crystals, where the propagation constants of two polarizations of light display a Dirac point degeneracy~\cite{hamilton1837}. While this curious phenomenon was revisited only occasionally in following century and a half~\cite{berry_conical}, interest in conical intersections rapidly expanded after the isolation of graphene in 2004 and discovery of many fascinating properties connected to its Dirac cones~\cite{graphene_review}. This has inspired the emulation of graphene in other fields including photonics. Consequently, most demonstrated examples of conical intersections to date are based on graphene's honeycomb lattice structure, since its symmetries guarantee the existence of Dirac cones, even in weak lattices for which the tight binding approximation is no longer valid~\cite{math_paper}. 

\begin{figure}

\includegraphics[width=\columnwidth]{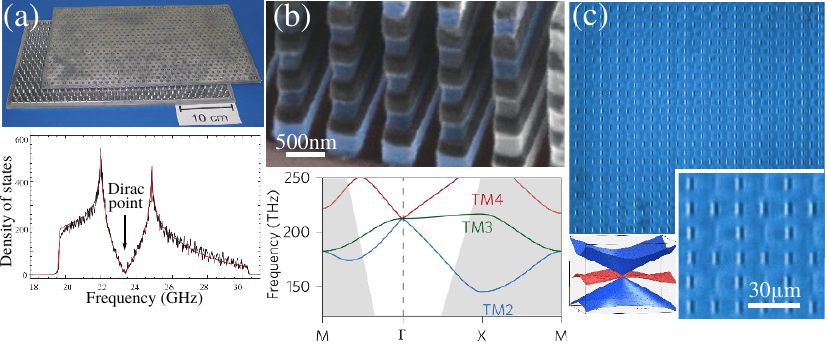}

\caption{Conical intersections in photonic systems from microwave to optical frequencies. (a) A triangular lattice formed by superconducting microwave resonator, displaying half-integer pseudospin Dirac cones. The measured density of states vanishes linearly as the Dirac point frequency is approached. Adapated from Ref.~\cite{dietz2013}. (b) All-dielectric zero index infrared metamaterial with band structure displaying conical intersection at $\Gamma$ point, inside the free-space light cone depicted in white (adapted from Ref.~\cite{moitra2013}). (c) Femtosecond laser-written Lieb lattice in silica glass, with integer pseudospin conical intersection at Brillouin zone corner (courtesy Falko Diebel, M\"unster University).}

\label{fig:examples}

\end{figure}

The honeycomb or hexagonal lattice geometry has now been studied theoretically and experimentally in many different photonic systems including metamaterials and photonic crystals~\cite{haldane2008,onoda2009,jukic2012}, plasmonic nanoparticles~\cite{yannopapas2011, weick2013}, photonic lattices~\cite{peleg2007,rechtsman2013,song2015}, and microwave resonator arrays~\cite{bittner2010,kuhl2010,bellec2013,dietz2013}, eg. Fig.~\ref{fig:examples}(a). There are also similar studies in the closely related settings of cold atoms in optical lattices~\cite{zhu2007,tarruell2012}, exciton-polariton condensates~\cite{jacqmin2014}, and optomechanical arrays~\cite{schmidt2015}. Arguably our understanding of wave dynamics at the Dirac cones in the linear, lossless limit is now mature; open topics of research involve the interplay of this linear band structure with subtleties peculiar to each system, such as driven-dissipative dynamics, quantum phenomena, and the role of interactions or nonlinearities. Detailed discussion of these effects in artificial honeycomb lattices can be found in Ref.~\cite{polini2013}.

A second breakthrough in the engineering of systems hosting conical intersections was the realization that the specific symmetries of the honeycomb lattice are not essential; conical intersections can also be created and moved through the Brillouin zone as \emph{accidental} degeneracies via the tuning of system parameters~\cite{asano2011,dirac_review}. This approach is especially useful for metamaterials and photonic crystals, as it opens up new lattice geometries hosting conical intersections for beams at normal incidence~\cite{liu2011,yannopapas2011,sakoda2012} and intersections based on degeneracies of transverse electric (TE) and transverse magnetic (TM) modes~\cite{khanikaev2012}. Typically the degeneracy is achieved by tuning parameters such as dielectric rod radius or filling factor in photonic crystals~\cite{huang2011,mei2012,chua2014}, refractive index contrast of nanoparticles~\cite{yannopapas2011}, lattice period~\cite{zhang2015}, or coupling strength between resonators~\cite{dirac_review,chong2015}.

Curiously in many cases, especially photonic metamaterials~\cite{huang2011,moitra2013,on_chip}, the accidental degeneracy also involves a third, auxiliary flat band, see Fig.~\ref{fig:examples}(b). For example in systems with $T$ ``time reversal'' symmetry this is required for intersections at the Brillouin zone centre ($\Gamma$ point; invariant under $T$), since Dirac cones must appear in pairs that are swapped by $T$. Although these cones are commonly (and inaccurately) called Dirac cones in the literature, this name overlooks some fundamental differences.

In fact, the auxiliary flat band is a signature of a higher order or ``bosonic'' conical intersection, which has distinctly different properties~\cite{fang2015}. Namely, wave propagation is governed by an \emph{integer} pseudospin variant of the Dirac equation, which displays unique effects such as resonant all-angle Klein tunnelling through potential barriers~\cite{shen2010,urban2011}. In this case, the prototypical example for studying the properties of integer pseudospin intersections has been the square-like ``Lieb lattice'' shown in Fig.~\ref{fig:examples}(c), originally proposed for cold atoms~\cite{shen2010,apaja2010,goldman2011,taie2015} and recently realized as a photonic lattice~\cite{leykam2012,guzman-silva2014,vicencio2015,mukherjee2015,diebel2015}.

Going further, there is now growing interest in studying and designing systems hosting generalized conical intersections with \emph{arbitrary} pseudospin $s$. Proposed designs include quasi-2D multilayer structures~\cite{dora2011}, or genuinely 2D optical lattices for multicomponent cold atoms in which laser-assisted tunnelling could be used to engineer the required matrix-valued hopping elements for arbitrary $s$~\cite{lan2011}. In photonics the design of such lattices remains a challenging open problem, demanding either increasingly complex lattice geometries or a way to engineer similar matrix-valued coupling terms between several near-degenerate internal modes. Perhaps a first step towards achieving this would be to demonstrate these generalized intersections in a lower dimensional (quasi-1D) setting, where for example analogues of the Dirac cone and $s=1$ cone have been realized~\cite{zeuner2012,mukherjee2015b}.

\section{Theory}
\label{sec:theory}

\begin{figure}

\includegraphics[width=\columnwidth]{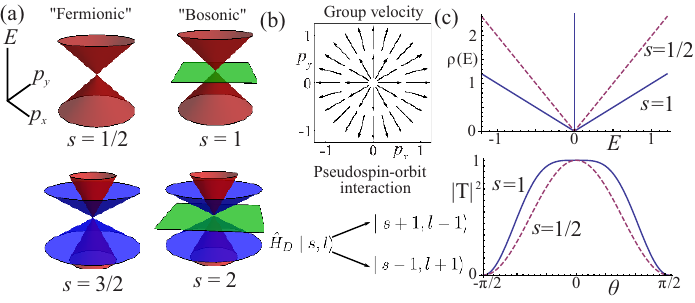}

\caption{(a) Conical intersections of different orders with pseudospin $s$. (b) Properties common to both ``fermionic'' (half-integer $s$) and ``bosonic'' (integer $s$) intersections. Top: group velocity field, directed radially with constant magnitude, singular at $\bs{p} = 0$. Bottom: time evolution $\hat{H}_D$ generating mixing between pseudospin $s$ and orbital $l$ degrees of freedom, conserving the total angular momentum $j = s + l$. (c) Differences between fermionic and bosonic intersections. Top: Density of states $\rho(E)$, with the bosonic intersection displaying a $\delta$ function singularity at $E = 0$. Bottom: Amplitude $|T|^2$ for Klein tunnelling through a potential step as a function of incident angle $\theta$, displaying a slower decay $|T|^2 \sim \cos^4 \theta$ in the bosonic case.}

\label{fig:theory}

\end{figure}

The disparate systems described above share universal long (transverse) wavelength dynamics at their conical intersections, with a Bloch wave spectrum in the lossless conservative limit obtained from the Hermitian eigenvalue problem
\begin{equation}
E_n (\bs{k} ) \mid u_n (\bs{k} ) \rangle = \hat{H} ( \bs{k} ) \mid u_n (\bs{k}) \rangle,
\end{equation}
where $E_n$ is the wave energy or frequency in the $n$th band, $\bs{k} = (k_x,k_y,0)$ is the in-plane wavevector, $\mid u_n (\bs{k}) \rangle$ is the Bloch function, and $\hat{H}(\bs{k})$ is an effective Hamiltonian. By applying the $\bs{k} \cdot \bs{p}$ perturbation theory, one can rigorously show that if the spectrum $E_n(\bs{k})$ displays a degeneracy at a point in the Brillouin zone $\bs{K}$, the dispersion plotted in Fig.~\ref{fig:theory}(a) is locally linear and remarkably described by the relativistic Dirac-Weyl Hamiltonian~\cite{mei2012},
\begin{equation}
\hat{H}_D(\bs{p}) = v_0 (\boldsymbol{p} \cdot \boldsymbol{\hat{S}}), \label{eq:dirac}
\end{equation}
where the energy and wavevector $\bs{p} \equiv \bs{k} - \bs{K}$ are measured with respect to the degeneracy, cone angle $v_0$ acts as the effective speed of light, and $\bs{\hat{S}} = (\hat{S}_x,\hat{S}_y,\hat{S}_z)$ are spin operators for a particle with spin $s$, which form a ``pseudospin''.

This pseudospin can be defined in terms of the symmetries of the degenerate Bloch modes at $\bs{p} = 0$~\cite{sakoda2012c,leykam_thesis}. It forms an analogue of ``real'' spin describing how the Bloch waves are distributed between near-degenerate internal or ``microscopic'' states of the periodic medium, such as the two inequivalent sublattices of a honeycomb lattice, different resonances of meta-atoms (monopolar/dipolar), or layers of a multilayer structure (eg. bilayer graphene). The key property of a conical intersection is the nontrivial coupling of this \emph{pseudo}spin to the \emph{real} momentum $\bs{p}$, which signifies a failure of a single band approximation and the necessity of a multi-band model. Nevertheless, complex multi-band effects can be understood intuitively via analogies with real spin. 

For example, a pseudospin-orbit coupling is revealed by introducing pseudospin raising/lowering operators $\hat{S}_{\pm} = \hat{S}_x \pm i \hat{S}_y$ and converting Eq.~\eqref{eq:dirac} to polar coordinates $\bs{p} = (p \cos \theta, p \sin \theta)$,
\begin{equation}
\hat{H}_D(\bs{p}) = v_0 p \left( e^{-i \theta} \hat{S}_+ + e^{i \theta} \hat{S}_- \right).
\end{equation}
Thus, wave dynamics at conical intersections generically involve conversion between different pseudospin states, and the generation of phase singularities $e^{\pm i \theta}$ (quantized vortices), demonstrating a deep connection between microscopic pseudospin and macroscopic angular momentum that is also directly observable, eg. in the optical absorption of graphene~\cite{mecklenburg2011,trushin2011} or conical diffraction experiments~\cite{song2015,diebel2015}.

Other features of the Dirac-Weyl Hamiltonian Eq.~\eqref{eq:dirac} can be derived using the basic properties of the pseudospin operators~\cite{leykam_thesis}. Energy eigenvalues are isotropic with linear dispersion $E_n(\bs{p}) = n v_0 p$ and constant group velocity $\nabla_{\bs{p}} E = n v_0 \bs{p} / p$ plotted in Fig.~\ref{fig:theory}(b), implying a linearly vanishing density of states $\rho ( E ) \propto |E|/n$ in contrast to the constant density of states associated with ordinary parabolic dispersion. Here the band index $n = -s,-s+1,...,s$ reveals a fundamental difference between ``bosonic'' integer and ``fermionic'' half-integer pseudospins $s$, namely the appearance of a zero energy flat band with divergent density of states in the former. Furthermore, the $2\pi$ Berry phase of the bosonic intersection leads to perfect Klein tunnelling of waves through potential steps across a wider range of incident angles compared to the Dirac cone displaying perfect transmission only at normal incidence~\cite{shen2010,urban2011}. These differences are summarized in Fig.~\ref{fig:theory}(c).

The two classes of intersections respond very differently to perturbations. For example, the Pauli operators $\hat{\sigma}_j$ associated with $s=1/2$ square to the identity, $\hat{\sigma}_j^2 = \hat{1}$, implying that the only way to open a gap at a Dirac cone while preserving the spatial symmetries of the lattice is to break the microscopic symmetry between the two pseudospin eigenstates, eg. by detuning two sublattices in the honeycomb lattice. This ``protection'' can be understood in terms of the quantized $\pi$ Berry phase accumulated when encircling a Dirac cone~\cite{dirac_review}. On the other hand, this property no longer holds for higher values of $s$: at a bosonic $s=1$ intersection with Berry phase 0 (mod $2\pi$)~\cite{photonic_dirac_cone}, symmetry only protects the degeneracy of two modes (eg. orthogonal dipolar modes in a metamaterial), while the third (typically a monopole mode) must be tuned to achieve an accidental degeneracy and conical intersection.

As noted above, in systems with time reversal ($T$) symmetry, Dirac cones appear in pairs related via $T$ symmetry, a consequence of their ``fermionic'' $\pi$ Berry phase. Hence in contrast to their ``bosonic'' $s=1$ counterparts, they cannot appear at $T$-invariant points of the Brillouin zone such as the $\Gamma$ point (required for device operation in the ``metamaterial'' regime). The only way around this restriction is to break $T$, eg. by considering magnetic optical media~\cite{lu2016}, or by modulating the system along the propagation direction~\cite{leykam2016}.

These two simplest examples, $s=1/2$ and $s=1$, are already sufficient to reveal the fundamental differences between ``fermionic'' and ``bosonic'' intersections. We note that for higher values of $s$, one must distinguish between genuine higher representations of $s$, and multiple degenerate copies of a lower $s$, for example the ``double Dirac cones'' of Refs.~\cite{sakoda2012b,li2015} employing accidental degeneracy between pairs of dipole and quadrupole modes to create two copies of the $s=1/2$ cone.

\section{Applications}
\label{sec:applications}

The unconventional, ``relativistic'' properties of conical intersections open up a variety of photonic applications, which typically exploit either their linear dispersion and density of states, or the nontrivial winding and Berry phase of the eigenmodes around the intersection. 

\begin{figure}

\includegraphics[width=\columnwidth]{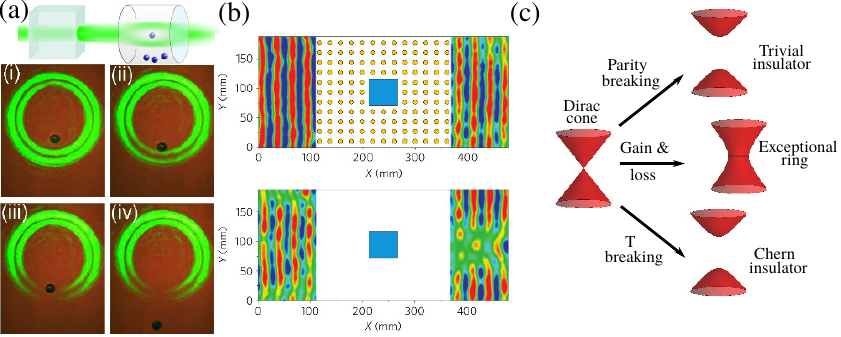}

\caption{Some examples of applications of conical intersections. (a) Experimental images from Ref.~\cite{conical_trapping_2} of the trapping and unloading of a $40\mu$m glass sphere (black circle) from a conical diffraction-based three-dimensional bottle beam, with transverse profile consisting of two bright rings separated by a dark ring. (b) Experimental observation from Ref.~\cite{huang2011} of cloaking in a zero index microwave metamaterial formed by a pseudospin 1 conical intersection. Top: an incident plane wave passes through the zero-index metamaterial with embedded metallic scatterer (blue square) without disruption. Bottom: for comparison, in the absence of the metamaterial the scatterer produces a shadow. (c) Novel phases accessible by weakly perturbing a pseudospin 1/2 conical intersection with different symmetry breaking terms.}
\label{fig:applications}

\end{figure}

The first predicted effect of a conical intersection was the formation of characteristic double ring diffraction pattern called conical diffraction~\cite{hamilton1837,berry_conical}, which is caused by the singular group velocity plotted in Fig.~\ref{fig:theory}(b): all wavevectors of a Gaussian beam have a group velocity directed radially outwards, with constant magnitude. Thus propagation leaves a dark central core surrounded by rings of constant thickness, which is useful for the optical trapping of particles. Applications span from the manipulation of biological samples to probing the physics of condensates of ultracold atoms~\cite{tweezers,conical_trapping_3}. A variety of trapping configurations are possible, such as using the dark core as a 3D bottle beam, eg. Fig.~\ref{fig:applications}(a). Alternatively in combination with a confining potential in the longitudinal (propagation) direction, one can create quasi-1D ring traps using either the two bright rings or the separating dark ring~\cite{conical_trapping_1,conical_trapping_2}. Additionally, the conversion between different pseudospin states during conical diffraction enables the controlled generation of orbital angular momentum, optical vortices, and other wave singularities~\cite{berry2005,dennis2009,song2015,diebel2015}. 

A linearly vanishing density of states as in Figs.~\ref{fig:examples}(a) and Fig.~\ref{fig:theory}(c) is a useful property for the design of photonic crystal slab lasers. The vanishing density of states promotes single mode lasing over a wide modal area when lasing occurs at a Dirac point, leading to more favourable scaling of the Purcell factor with the lasing area~\cite{bravo-abad2012}. These useful properties can even persist for the integer pseudospin intersections, because typically one of the conical bands and the flat band (with divergent density of states) couple more strongly into radiative modes, leaving a single conical band with higher quality factor $Q$ and a lower lasing threshold~\cite{chua2014}. Conversely, if the output coupling is designed such that the flat band has the highest $Q$, its strong degeneracy and divergent density of states in Fig.~\ref{fig:theory}(c) should lead to strongly multimode lasing with no long range phase order~\cite{frustrated_laser}. 

The flat dispersion bands associated with the bosonic integer pseudospin intersections have other interesting applications beyond lasing. Their vanishing group velocity and strong degeneracy can be directly observed via the existence of strongly localized yet propagation invariant (nondiffracting) wavepackets~\cite{vicencio2015,mukherjee2015,mukherjee2015b}. This nondiffracting behavior leads to a strong sensitivity to further perturbations such as disorder or interactions~\cite{flach2014,apaja2010,taie2015}, analogous to slow light-enhanced nonlinear optical effects. Flat bands can also be generated by pairs of fermionic Dirac cones, where they appear as bands of surface states linking pairs of cones~\cite{jukic2012,rechtsman2013}.

Conical intersections enable the realization of superior zero-index metamaterials. First generation designs were limited to vanishing permittivity $\epsilon$ with permeability $\mu \sim 1$. While providing $n = \sqrt{\epsilon \mu} \approx 0$, the large ratio $\mu / \epsilon$ leads to strong impedance mismatch with free space modes, limiting applications. More useful designs require $\mu \approx \epsilon \approx 0$ for impedance matching. In fact, this simultaneous vanishing of $\mu$ and $\epsilon$ implies a locally linear dispersion relation, namely a conical intersection. This has lead to new designs for low loss, all-dielectric zero index metamaterials based on accidentally degenerate, pseudospin 1 conical intersections at microwave~\cite{huang2011} and infrared frequencies~\cite{moitra2013,on_chip}. Among the numerous and interdisciplinary applications of these zero index materials are supercollimation~\cite{fang2015}, leaky-wave antennas~\cite{memarian2015}, and cloaking, illustrated in Fig.~\ref{fig:applications}(b). Notably, a pseudospin 1 intersection is not strictly necessary to realize a zero index metamaterial; alternate designs based on doubly degenerate Dirac cones have also been proposed~\cite{sakoda2012b,li2015}, but not yet realized in experiment.

On a more fundamental level, the realization of conical intersections in photonic and matter wave systems allows the direct observation of relativistic quantum analogies not easily visible in systems such as graphene~\cite{analogies}, for example the parity anomaly~\cite{schomerus2013}, along with the study of surface effects such as edge states that are notoriously hard to control in condensed matter systems~\cite{onoda2009,kuhl2010,jukic2012,rechtsman2013}. When a conical dispersion is restricted to a finite system, the intersection is replaced by a (small) band gap, due to the vanishing density of states. In contrast to regular Bragg gaps, the wave transmission $T \sim 1/L$ scales pseudo-diffusively with the lattice size $L$, for both ``fermionic'' and ``bosonic'' intersections~\cite{sepkhanov2007,diffusive_experiment,yannopapas2011,transmission_spin1}. On the other hand, the \emph{reflected} beam is sensitive to the type of intersection, displaying a suppression of coherent backscattering (weak antilocalization) due to the $\pi$ Berry phase of ``fermionic'' intersections~\cite{cbs}. 

Finally, the point degeneracy between multiple spectral bands is itself extremely useful, enabling the creation of other novel dispersion relations under modest symmetry-breaking perturbations, eg. Fig.~\ref{fig:applications}(c). For example, at a Dirac point the typically weak magneto-optic effect is still strong enough to open a ``topologically nontrivial'' band gap to realize a photonic topological insulator supporting unidirectional edge states~\cite{haldane2008}. For further discussion on this topic we recommend the recent review article on topological photonics, Ref.~\cite{lu2014}. Under balanced gain and loss, the real conical spectrum becomes complex, emulating superluminal ``tachyonic'' dispersion~\cite{szameit2011,chong2015}. This superluminal dispersion is associated with the appearance of the non-Hermitian analogue of a conical intersection, ``exceptional point'' degeneracies~\cite{exceptional_point}, which form generically from \emph{any} accidental conical intersection when open boundary conditions are applied~\cite{zhen2015}. Importantly, all these exotic effects can also be understood within relatively simple coupled mode models describing the near-degenerate modes at conical intersections.

\section{Outlook}
\label{sec:conclusion}

We have seen significant progress in understanding the nature and ubiquity of conical intersections in two-dimensional periodic media, with interest now directed towards studying applications that go beyond demonstrating simple analogies with the electronic structure of graphene. Of these, the realization of zero index metamaterials is perhaps the most prominent~\cite{huang2011,moitra2013,on_chip}. Nevertheless there is still the potential for further cross-fertilization between different fields, going beyond the ``artificial graphene'' paradigm. A striking example of this is the recent discovery that a periodic lattice is not strictly necessary; conical intersections can also occur in quasiperiodic lattices~\cite{quasicrystal_conical}.

The idea of generating conical intersections via accidental degeneracies, which is especially useful for strongly coupled systems in which the tight binding approximation is often invalid, has not yet significantly filtered through to the matter wave and cold atom literature. Likely reasons for this are the need to precisely engineer the potential landscape to hit the accidental degeneracy in the former, and the extremely good validity of the tight binding approximation in the latter when deep optical lattices are used. However, in systems such as exciton-polariton condensates where the energy resolution is typically a limiting factor, accidental degeneracies may be superior to tight binding limit conical intersections in certain applications (recall that the energy bandwidth vanishes as the tight binding limit is approached). Thus this is an avenue that deserves further attention in the context of recent advances in generating structured potentials for exciton-polaritons~\cite{polariton_trapping}. Perhaps because of this bandwidth limitation, a ``bosonic'' conical intersection has not yet been realized for exciton-polariton condensates.

Conversely, the study of interactions and nonlinearities which are important for Bose-Einstein condensates~\cite{nonlinear_dirac_1,nonlinear_dirac_2} and cold atoms has not been significantly pursued in the photonic domain, with only a handful theoretical studies based on experimentally feasible parameters published so far, for example demonstrating bistability~\cite{bistability} and ability for phase-matched second harmonic generation based on the vanishing effective refractive index~\cite{shg}. It is therefore of interest to fabricate photonic media combining strong optical nonlinearities with conical intersections to enable the study of both all-optical beam control, and fundamental effects such as the existence and stability of solitons~\cite{peleg2007,nonlinear_dirac_2,cuevas2015}.

Finally, we stress that these ideas are not just limited to photonic and matter waves. Beyond the scope of this review, other areas where conical intersections are attracting interest include the surface states of topological insulators~\cite{topo_review} and three-dimensional topological photonic crystals~\cite{lu2016}, and the bulk dispersion of phononic metamaterials~\cite{mei2012,liu2011,liang2013}. One interesting potential application of the latter is protecting structures from earthquakes using a seismic wave cloak~\cite{seismic}.

\section*{Acknowledgement}

This research was supported by the Singapore National Research Foundation under grant No. NRFF2012-02.


\begin{thebibliography}{99}

\bibitem{condensed_review}
D. Xiao, M.-C. Chang, and Q. Niu, {\it Berry phase effects on electronic properties}, Rev. Mod. Phys. {\bf 82}, 1959 (2010).

\bibitem{jahn-teller}
H.~A. Jahn and E. Teller, {\it Stability of polyatomic moleculares in degenerate electronic states. I. Orbital degeneracy}, Proc. Royal Soc. A {\bf 161}, 220 (1937).

\bibitem{chemical_review}
G.~A. Worth and L.~S. Cederbaum, {\it Beyond Born-Oppenheimer: Molecular dynamics through a conical intersection}, Ann. Rev. Phys. Chem. {\bf 55}, 127 (2004).

\bibitem{graphene_review}
A.~K. Geim and K.~S. Novoselov, {\it The rise of graphene}, Nature Mater. {\bf 6}, 183 (2007).
%

\bibitem{dennis2009}
M.~R. Dennis, K. O'Holleran, and M.~J. Padgett, {\it Singular optics: Optical vortices and polarization singularities}, Prog. Opt. {\bf 53}, 293 (2009).

\bibitem{lu2014}
L. Lu, J.~D. Joannopoulos, and M. Solja\v{c}i\'{c}, {\it Topological photonics}, Nature Photon. {\bf 8}, 821 (2014).

\bibitem{bec1}
F. Dalfovo, S. Giorgino, L.~P. Pitaevskii, and S. Stringari, {\it Theory of Bose-Einstein condensation in trapped gases}, Rev. Mod. Phys. {\bf 71}, 463 (1999).

\bibitem{bec2}
T. Byrnes, N.~Y. Kim, and Y. Yamamoto, {\it Exciton-polariton condensates}, Nature Phys. {\bf 10}, 803 (2014).

\bibitem{klein_tunnelling}
M.~I. Katsnelson, K.~S. Novoselov, and A.~K. Geim, {\it Chiral tunnelling and the Klein paradox in graphene}, Nature Phys. {\bf 2}, 620 (2006).

\bibitem{on_chip}
Y. Li, S. Kita, P. Munoz, O. Reshef, D.~I. Vulis, M. Yin, M. Loncar, and E. Mazur, {\it On-chip zero-index metamaterials}, Nature Photon. {\bf 9}, 738 (2015).


\bibitem{polini2013}
M. Polini, F. Guinea, M. Lewenstein, H.~C. Manoharan, and V. Mellegrini, {\it Artificial honeycomb lattices for electrons, atoms and photons}, Nature Nanotech. {\bf 8}, 625 (2013).
%



\bibitem{hamilton1837}
W.~R. Hamilton, {\it Third supplement to an essay on the theory of systems of rays}, Trans. Royal Irish Acad. {\bf 17}, 1 (1837).
%
\bibitem{berry_conical}
M.~V. Berry and M.~R. Jeffrey, {\it Conical diffraction: Hamilton's diabolical point at the heart of crystal optics}, Progress in Optics {\bf 50}, 13 (2007).
%

\bibitem{math_paper}
C. L. Fefferman and M.~I. Weinstein, {\it Honeycomb lattice potentials and Dirac points}, J. Amer. Math. Soc. {\bf 25}, 1169 (2012).



\bibitem{haldane2008}
F.~D.M. Haldane and S. Raghu, {\it Possible realization of directional optical waveguides in photonic crystals with broken time-reversal symmetry}, Phys. Rev. Lett. {\bf 100}, 013904 (2008).

\bibitem{onoda2009}
T. Ochiai and M. Onodo, {\it Photonic analog of graphene model and its extension: Dirac cone, symmetry, and edge states}, Phys. Rev. B {\bf 80}, 155103 (2009).

\bibitem{jukic2012}
D. Jukic, H. Buljan, D.-H. Lee, J.~D. Joannopoulos, and M. Solja\v{c}i\'{c}, {\it Flat photonic surface bands pinned between Dirac points}, Opt. Lett. {\bf 37}, 5262 (2012).
%

\bibitem{yannopapas2011}
V. Yannopapas and A. Vanakaras, {\it Dirac point in the photon dispersion relation of a negative/zero/positive-index plasmonic metamaterial}, Phys. Rev. B {\bf 84}, 045128 (2011).
%
\bibitem{weick2013}
G. Weick, C. Woollacott, W.~L. Barnes, O. Hess, and E. Mariani, {\it Dirac-like plasmons in honeycomb lattices of metallic nanoparticles}, Phys. Rev. Lett. {\bf 110}, 106801 (2013).
%

\bibitem{peleg2007}
O. Peleg, G. Bartal, B. Freedman, O. Manela, M. Segev, and D.~N. Christodoulides, {\it Conical diffraction and gap solitons in honeycomb lattices}, Phys. Rev. Lett. {\bf 98}, 103901 (2007).
%

\bibitem{rechtsman2013}
M.~C. Rechtsman, Y. Plotnik, J.~M. Zeuner, D. Song, Z. Chen, A. Szameit, and M. Segev, {\it Topological creation and destruction of edge states in photonic graphene}, Phys. Rev. Lett. {\bf 111}, 103901 (2013); Y. Plotnik, M.~C. Rechtsman, D. Song, M. Heinrich, J.~M. Zeuner, S. Nolte, Y. Lumer, N. Malkova, J. Xu, A. Szameit, Z. Chen, and M. Segev, {\it Observation of unconventional edge states in photonic graphene}, Nature. Mater. {\bf 13}, 57 (2014).
%

\bibitem{song2015}
D. Song, V. Paltoglou, S. Liu, Y. Zhu, D. Gallardo, L. Tang, J. Xu, M. Ablowitz, N.~K. Efremidis, and Z. Chen, {\it Unveiling pseudospin and angular momentum in photonic graphene}, Nature. Comms. {\bf 6}, 6272 (2015).
%

\bibitem{bittner2010}
S. Bittner, B. Dietz, M. Miski-Oglu, P. Oria Iriarte, A. Richter, and F. Sch\"afer, {\it Observation of a Dirac point in microwave experiments with a photonic crystal modeling graphene}, Phys. Rev. B {\bf 82}, 014301 (2010).
%

\bibitem{kuhl2010}
U. Kuhl, S. Barkhofen, T. Tudorovskiy, H.-J. St\"ockmann, T. Hossain, L. de Forges de Parny, and F. Mortessagne, {\it Dirac point and edge states in a microwave realization of tight-binding graphene-like structures}, Phys. Rev. B {\bf 82}, 094308 (2010).
%

\bibitem{bellec2013}
M. Bellec, U. Kuhl, G. Montambaux, and F. Mortessagne, {\it Tight binding couplings in microwave artificial graphene}, Phys. Rev. B {\bf 88}, 115437 (2013).
%
\bibitem{dietz2013}
B. Dietz, F. Iachello, M.Miski-Oglu, N. Pietralla, A. richter, L. von Smekal, and J. Wambach, {\it Lifshitz and excited-state quantum phase transitions in microwave Dirac billiards}, Phys. Rev. B {\bf 88}, 104101 (2013).


\bibitem{zhu2007}
S.-L. Zhu, B. Wang, and L.-M. Duan, {\it Simulation and detection of Dirac fermions with cold atoms in an optical lattice}, Phys. Rev. Lett. {\bf 98}, 260402 (2007).

\bibitem{tarruell2012}
L. Tarruell, D. Greif, T. Uehlinger, G. Jotzu, and T. Esslinger, {\it Creating, moving and merging Dirac points with a Fermi gas in a tunable honeycomb lattice} Nature {\bf 483}, 302 (2012).
%
\bibitem{jacqmin2014}
T. Jacqmin, I. Carusotto, I. Sagnes, M. Abbarchi, D.~ D. Solnyshkov, G. Malpuech, E. Galopin, A. Lemaitre, J. Bloch, and A. Amo, {\it Direct observation of Dirac cones and a flatband in a honeycomb lattice for polaritons}, Phys. Rev. Lett. {\bf 112}, 116402 (2014).
%
\bibitem{schmidt2015}
M. Schmidt, V. Peano, and F. Marquardt, {\it Optomechanical Dirac physics}, New. J. Phys. {\bf 17}, 023025 (2015).
%




\bibitem{dirac_review}
M. O. Goerbig, G. Montambaux, {\it Dirac fermions in condensed matter and beyond}, Seminaire Poincare {\bf 17}, 1 (2013).
%
\bibitem{asano2011}
K. Asano and C. Hotta, {\it Designing Dirac points in two-dimensional lattices}, Phys. Rev. B {\bf 83}, 245125 (2011).
%

\bibitem{liu2011}
F. Liu, Y. Lai, X. Huang, and C.~T. Chan, {\it Dirac cones at k=0 in phononic crystals} Phys. Rev. B {\bf 84}, 224113  (2011).
%
\bibitem{sakoda2012}
K. Sakoda, {\it Dirac cone in two- and three-dimensional metamaterials}, Opt. Exp. {\bf 20}, 3898 (2012).
%
\bibitem{khanikaev2012}
A.~B. Khanikaev, S.~H. Nousavi, W.-K. Tse, M. Kargarian, A.~H. MacDonald, and G. Shvets, {\it Photonic topological insulators}, Nature Mater. {\bf 12}, 233 (2013).
%




\bibitem{huang2011}
X. Huang, Y. Lai, Z.~H. Hang, H. Zheng, and C.~T. Chan, {\it Dirac cones induced by accidental degeneracy in photonic crystals and zero-refractive-index materials}, Nature Mater. {\bf 10}, 582 (2011).
%

\bibitem{mei2012}
J. Mei, Y. Wu, C.~T. Chan, and Z.-Q. Zhang, {\it First-principles study of Dirac and Dirac-like cones in phononic and photonic crystals}, Phys. Rev. B {\bf 86}, 035141 (2012).
%

\bibitem{chua2014}
S.-L. Chua, L. Lu, J. Bravo-Abad, J.~D. Joannopoulos, and M. Solja\v{c}i\'{c}, {\it Larger-area single-mode photonic crystal surface-emitting lasers enabled by an accidental Dirac point}, Opt. Lett. {\bf 39}, 2072 (2014).

\bibitem{zhang2015}
P. Zhang, C. Fietz, P. Tassin, T. Koschny, and C.~M. Soukoulis, {\it Numerical investigation of the flat band Bloch modes in a 2D photonic crystal with Dirac cones}, Opt. Express {\bf 23}, 10444 (2015).
%
\bibitem{chong2015}
Y.~D. Chong and M.~C. Rechtsman, {\it Tachyonic dispersion in coherent networks}, J. Opt. {\bf 18}, 014001 (2016).
%

\bibitem{moitra2013}
P. Moitra, Y. Yang, Z. Anderson, I.~I. Kravchenko, D.~P. Briggs, and J. Valentine, {\it Realization of an all-dielectric zero-index optical metamaterial}, Nature Photon. {\bf 7}, 791 (2013). 
%

\bibitem{fang2015}
A. Fang, Z.~Q. Zhang, S.~G. Louie, and C.~T. Chan, {\it Klein tunneling and supercollimation of pseudospin-1 electromagnetic waves}, Phys. Rev. B {\bf 93}, 035422 (2016).
%

\bibitem{shen2010}
R. Shen, L.~B. Shao, B. Wang, and D.~Y. Xing, {\it Single Dirac cone with a flat band touching on line-centred-square optical lattices}, Phys. Rev. B {\bf 81}, 041410 (2010).
%
\bibitem{urban2011}
D.~F. Urban, D. Bercioux, M. Wimmer, and W. H\"ausler, {\it Barrier transmission of Dirac-like pseudospin-one particles}, Phys. Rev. B {\bf 84}, 115136 (2011).
%

\bibitem{apaja2010}
V. Apaja, M. Hyrk\"as, and M. Manninen, {\it Flat bands, Dirac cones and atom dynamics in an optical lattice}, Phys. Rev. A {\bf 82}, 041402 (2010).
%

\bibitem{goldman2011}
N. Goldman, D.~F. Urban, and D. Bercioux, {\it Topological phases for fermionic cold atoms on the Lieb lattice}, Phys. Rev. A {\bf 83}, 063601 (2011).
%

\bibitem{taie2015}
S. Taie, H. Ozawa, T. Ichinose, T. Nishio, S. Nakajima, and Y. Takahashi, {\it Coherent driving and freezing of bosonic matter wave in an optical Lieb lattice}, Science Advances {\bf 1}, e1500854 (2015).



\bibitem{leykam2012}
D. Leykam, O. Bahat-Treidel, and A.~S. Desyatnikov, {\it Pseudospin and nonlinear conical diffraction in Lieb lattices}, Phys. Rev. A {\bf 86}, 031805 (2012).
%

\bibitem{guzman-silva2014}
D. Guzman-Silva, C. Meija-Cortes, M.~A. Bandres, M.~C. Rechtsman, S. Weimann,
S. Nolte, M. Segev, A. Szameit, and R.~A. Vicencio, {\it Experimental observation of bulk and edge transport in photonic Lieb lattices}, New J. Phys. {\bf 16}, 063061 (2014).

\bibitem{diebel2015}
F. Diebel, D. Leykam, S. Kroesen, C. Denz, and A.~S. Desyatnikov, {\it Observation of conical diffraction in photonic Lieb lattices}, in Advanced Photonics, OSA Technical Digest (online), NW3A.1 (2014).

\bibitem{vicencio2015}
R. A. Vicencio, C. Cantillano, L. Morales-Inostroza, B. Real, C. Meijia-Cortes, S. Weimann, A. Szameit, and M. I. Molina, {\it Observation of localized states in Lieb photonic lattices}, Phys. Rev. Lett. {\bf 114}, 245503 (2015).
%

\bibitem{mukherjee2015}
S. Mukherjee, A. Spracklen, D. Choudhury, N. Goldman, P. \"Ohberg, E. Andersson, and R. R. Thomson, {\it Observation of a localized flat-band state in a photonic Lieb lattice}, Phys. Rev. Lett. {\bf 114}, 245504 (2015).
%



\bibitem{dora2011}
B. Dora, J. Kailasvuori, and R. Moessner, {\it Lattice generalization of the Dirac equation to general spin and the role of the flat band}, Phys. Rev. B {\bf 84}, 195422 (2011).


\bibitem{lan2011}
Z. Lan, N. Goldman, A. Bermudez, W. Lu, and P. \"Ohberg, {\it Dirac-Weyl fermions with arbitrary spin in two-dimensional optical superlattices}, Phys. Rev. B {\bf 84},
165115 (2011).
%

\bibitem{zeuner2012}
J.~M. Zeuner, N.~K. Efremidis, R. Keil, F. Dreisow, D.~N. Christodoulides, A. T\"unnermann, S. Nolte, and A. Szameit, {\it Optical analogues for massless Dirac particles and conical diffraction in one dimension}, Phys. Rev. Lett. {\bf 109}, 023602 (2012).

\bibitem{mukherjee2015b}
S. Mukherjee and R.~R. Thomson, {\it Observation of localized flat-band modes in a quasi-one-dimensional photonic rhombic lattice}, Opt. Lett. {\bf 40}, 5443 (2015).



\bibitem{sakoda2012c}
K. Sakoda, {\it Proof of the universality of mode symmetries in creating photonic Dirac cones}, Opt. Exp. {\bf 20}, 25181 (2012).


\bibitem{leykam_thesis}
D. Leykam, {\it Wave and spectral singularities in photonic lattices}, Ph.D. Thesis, The Australian National University, 2015.

\bibitem{mecklenburg2011}
M. Mecklenburg and B.~C. Regan, {\it Spin and the honeycomb lattice: Lessons from graphene}, Phys. Rev. Lett. {\bf 106}, 116803 (2011).
%

\bibitem{trushin2011}
M. Trushin and J. Schliemann, {\it Pseudospin in optical and transport properties of graphene}, Phys. Rev. Lett. {\bf 107}, 156801 (2011).
%

\bibitem{photonic_dirac_cone}
C.~T. Chan, Z.~H. Hang, and X. Huang, {\it Dirac dispersion in two-dimensional photonic crystals}, Advances in OptoElectronics {\bf 2012}, 313984 (2012).


\bibitem{lu2016}
L. Lu, C. Fang, L. Fu, S.~G. Johnson, J.~D. Joannopoulos, and M. Solja\v{c}i\'{c}, {\it Symmetry-protected topological photonic crystal in three dimensions}, Nature Phys. 3611 (2016).

\bibitem{leykam2016}
D. Leykam, M.~C. Rechtsman, and Y.~D. Chong, {\it Anomalous topological phases and unpaired Dirac cones in photonic Floquet topological insulators}, arXiv:1601.01764.


\bibitem{sakoda2012b}
K. Sakoda, {\it Double Dirac cones in triangular-lattice metamaterials}, Opt. Exp. {\bf 20}, 9925 (2012).

\bibitem{li2015}
Y. Li and J. Mei, {\it Double Dirac cones in two-dimensional photonic crystals}, Opt. Exp. {\b 23}, 12089 (2015).



\bibitem{tweezers}
C. McDonald, C. McDougall, E. Rafailov, and D. McGloin, {\it Characterizing conical refraction optical tweezers}, Opt. Lett. {\bf 39}, 6691 (2014).


\bibitem{conical_trapping_3}
A. Turpin, J. Polo, Yu. V. Loiko, J. K\"uber, F. Schmaltz, T. K. Kalkandjiev, V. Ahufinger, G. Birkl, and J. Mompart, {\it Blue-detuned optical ring trap for Bose-Einstein condensates based on conical refraction}, Opt. Express {\bf 23}, 1638 (2015).

\bibitem{conical_trapping_1}
D.~P. O'Dwyer, C.~F. Phelan, K.~E. Ballantine, Y.~P. Rakovich, J.~G. Lunney, and J.~F. Donegan, {\it Conical diffraction of linearly polarised light controls the angular position of a microscopic object }, Opt. Express {\bf 18}, 27319 (2010).

\bibitem{conical_trapping_2}
A. Turpin, V. Shvedov, C. Hnatovsky, Yu.~V. Loiko, J. Mompart, and W. Krolikowski, {\it Optical vault: A reconfigurable bottle beam based on conical refraction of light}, Opt. Express {\bf 21}, 26335 (2013). 

\bibitem{berry2005}
M.~V. Berry, M.~R. Jeffrey, and M. Mansuripur, {\it Orbital and spin angular momentum in conical diffraction}, J. Opt. A {\bf 7}, 685 (2005). 


\bibitem{bravo-abad2012}
J. Bravo-Abad, J.~D. Joannopoulos, and M. Solja\v{c}i\'{c}, {\it Enabling single mode behavior over large areas with photonic Dirac cones}, Proc. Natl. Acad. Sci. USA {\bf 109}, 9761 (2012).
%
\bibitem{frustrated_laser}
M. Nixon, E. Ronen, A.~A. Friesem, and N. Davidson, {\it Observing geometric frustration with thousands of coupled lasers}, Phys. Rev. Lett. {\bf 110}, 184102 (2013).



\bibitem{flach2014}
S. Flach, D. Leykam, J.~D. Bodyfelt, P. Matthies, and A. S. Desyatnikov, {\it Detangling flat bands into Fano lattices}, EPL {\bf 105}, 30001 (2014).


\bibitem{memarian2015}
M. Memarian and G.~V. Eleftheriades, {\it Dirac leaky-wave antennas for continuous beam scanning from photonic crystals}, Nature Comms. {\bf 6}, 5855 (2015).
%


\bibitem{analogies}
S. Longhi, {\it Classical simulation of relativistic quantum mechanics in periodic optical structures}, App. Phys. B {\bf 104}, 453 (2011).

\bibitem{schomerus2013}
H. Schomerus and N.~Y. Halpern, {\it Parity anomaly and Landau-level lasing in strained photonic honeycomb lattices}, Phys. Rev. Lett. {\bf 110}, 013903 (2013).





\bibitem{sepkhanov2007}
R.~A. Sepkhanov, Ya.~B. Bazaliy, and C.~W.~J. Beenakker, {\it Extremal transmission at the Dirac point of a photonic band structure}, Phys. Rev. A {\bf 75}, 063813 (2007).
%

\bibitem{diffusive_experiment}
S.~R. Zandbergen and M.~J.~A. de Dood, {\it Experimental observation of strong edge effects on the pseudodiffusive transport of light in photonic graphene}, Phys. Rev. Lett. {\bf 104}, 043903 (2010).


\bibitem{transmission_spin1}
X. Wang, H.~T. Jiang, C. Yan, F.~S. Deng, Y. Sun, Y.~H. Li., Y.~L. Shi, and H. Chen, {\it Transmission properties near Dirac-like point in two-dimensional dielectric photonic crystals}, EPL {\bf 108}, 14002 (2014).

\bibitem{cbs}
R.~A. Sepkhanov, A. Ossipov, and C.~W.~J. Beenakker, {\it Extinction of coherent backscattering by a disordered photonic crystal with a Dirac spectrum}, EPL {\bf 85}, 14005 (2009).

\bibitem{szameit2011}
A. Szameit, M.~C. Rechtsman, O. Bahat-Treidel, and M. Segev, {\it PT-symmetry in honeycomb photonic lattices}, Phys. Rev. A {\bf 84}, 021806(R) (2011).

\bibitem{exceptional_point}
T. Gao, E. Estrecho, K.~Y. Bliokh, T.~C.~H. Liew, M.~D. Fraser, S. Brodbeck, M. Kamp, C. Schneider, S. H\"ofling, Y. Yamamoto, F. Nori, Y.~S. Kivshr, A.~G. Truscott, R.~G. Dall, and E.~A. Ostrovskaya, {\it Observation of non-Hermitian degeneracies in a chaotic exciton-polariton billiard}, Nature {\bf 526}, 554 (2015).

\bibitem{zhen2015}
B. Zhen, C.~W. Hsu, Y. Igarashi, L. Lu, I. Kaminer, A. Pick, S.-L. Chua, J.~D. Joannopoulos, and M. Solja\v{c}i\'{c}, {\it Spawning rings of exceptional points of out Dirac cones}, Nature {\bf 525}, 354 (2015).



\bibitem{quasicrystal_conical}
J.-W. Dong, M.-L. Chang, X.-Q. Huang, Z.~H. Hang, Z.-C. Zhong, W.-J. Chen, Z.-Y. Huang, and C.~T. Chan, {\it Conical dispersion and effective zero refractive index in photonic quasicrystals}, Phys. Rev. Lett. {\bf 114}, 163901 (2015).


\bibitem{polariton_trapping}
C. Schneider, K. Winkler, M.~D. Fraser, M. Kamp, Y. Yamamoto, E.~A. Ostrovskaya, and S. H\"ofling, {\it Exciton-polariton trapping and potential landscape engineeering}, arXiv:1510.07540 (2015).


\bibitem{nonlinear_dirac_1}
M.~J. Ablowitz, S.~D. Nixon, and Y. Zhu, {\it Conical diffraction in honeycomb lattices}, Phys. Rev. A {\bf 79}, 053830 (2009).

\bibitem{nonlinear_dirac_2}
L.~H. Haddad and L.~D. Carr, {\it The nonlinear Dirac equation in Bose-Einstein condensates: superfluid fluctuations and emergent theories from relativistic linear stability equations}, New J. Phys. {\bf 17}, 093037 (2015).

\bibitem{cuevas2015}
J. Cuevas-Maraver, P.~G. Kevrekidis, A. Saxena, A. Comech, and R. Lan, {\it Stability of solitary waves and vortices in a 2D nonlinear Dirac model}, arXiv:1512.03973.

\bibitem{bistability}
N. Mattiucci, M.~J. Bloemer, and G. D'Aguanno, {\it All-optical bistability and switching near the Dirac point of a 2-D photonic crystal}, Opt. Express {\bf 21}, 11862 (2013).

\bibitem{shg}
N. Mattiucci, M.~J. Bloemer, and G. D'Aguanno, {\it Phase-matched second harmonic generation at the Dirac point of a 2-D photonic crystal}, Opt. Express {\bf 22}, 6381 (2014).

\bibitem{topo_review}
M.~Z. Hasan and C.~L. Kane, {\it Colloquium: Topological insulators}, Rev. Mod. Phys. {\bf 82}, 3045 (2010); X.-L. Qi and S.-C. Zhang, {\it Topological insulators and superconductors}, Rev. Mod. Phys. {\bf 83}, 1057 (2011).


\bibitem{liang2013}
Z. Liang, T. Feng, S. Lok, F. Liu, K.~B. Ng, C.~H. Chan, J. Wang, S. Han, S. Lee, and J. Li, {\it Space-coiling metamaterials with double negativity and conical dispersion}, Scientific Rep. {\bf 3}, 1614 (2013).

\bibitem{seismic}
S. Brule, E.~H. Javelaud, S. Enoch, and S. Guenneau, {\it Experiments on seismic metamaterials: Molding surface waves}, Phys. Rev. Lett. {\bf 112}, 133901 (2014).

\bibitem{elganainy2015}
R. El-Ganainy, J.~I. Dadap, and R.~M. Osgood, Jr., {\it Optical parametric amplification via non-Hermitian phase matching}, Opt. Lett. {\bf 40}, 5086 (2015).

\end{thebibliography}
\end{document}